\documentclass{moriond}

\usepackage{braket}
\usepackage{slashed}
\usepackage{amsmath}
\usepackage{mathtools}
\usepackage{dsfont}

\def\bea{\begin{eqnarray}}
\def\eea{\end{eqnarray}}

\newcommand{\ii}{\mathrm{i}}

\newcommand{\Photo}{\includegraphics[height=35mm]{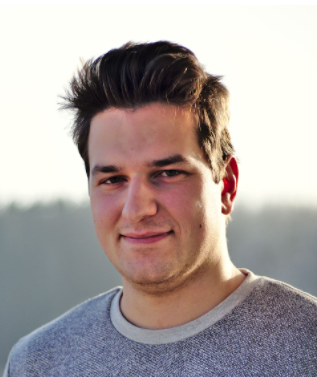}}

\begin{document}
\vspace*{4cm}
\title{UGR TESTS WITH ATOMIC CLOCKS AND ATOM INTERFEROMETERS
}

\author{F. DI PUMPO\\(on behalf of the QUANTUS and MIUnD collaborations)}

\address{Institut f{\"u}r Quantenphysik and Center for Integrated Quantum Science and Technology (IQ\textsuperscript{ST}), Universit{\"a}t Ulm, Albert-Einstein-Allee 11, D-89069 Ulm, Germany}

\maketitle\abstracts{
Atomic interference experiments test the universality of the coupling between matter-energy and gravity at different spacetime points, thus being in principle able to probe possible violations of the universality of the gravitational redshift (UGR).
In this contribution, we introduce a UGR violation model and then discuss UGR tests performed by atomic clocks and atom interferometers on the same footing. 
We present a large class of atom-interferometric geometries which are sensitive to violations of UGR.}

\section{Introduction}
Phases of matter waves can be connected\,\cite{Storey1994} to proper time\,\cite{Einstein1905,Einstein1911}.
Therefore, matter waves are in principle able to test the universality of the gravitational coupling to test bodies.
This universality is based on the Einstein equivalence principle (EEP) and translates into three basic assumptions\,\cite{Will2014,DiCasola2015}: local Lorentz invariance, universality of free fall (UFF), and local position invariance.
The last principle can be divided into the universality of gravitational redshift (UGR) and the universality of clock rates; we focus on UGR in this article.
EEP implies that gravity is a metric theory, demanding that every test body couples universally to gravity, independent of its composition.
UGR states that the proper-time difference between different heights in gravity passes in a composition-independent manner, and is usually tested\,\cite{Vessot1980,Chou2010,Herrmann2018,Delva2019,Takamoto2020,Bothwell2021} by atomic clocks\,\cite{Brewer2019,Oelker2019,Madjarov2019}.
On the other hand, UFF tests analyze whether gravitational accelerations of two test bodies are equal, and can be implemented via classical\,\cite{Touboul2019} or quantum test bodies, where the latter can for example be implemented with atom interferometers\,\cite{Schlippert2014,Asenbam2020}.
Consequently, the question\,\cite{Mueller2010,Wolf2010,Mueller2010b,Unnikrishnan2011,Schleich2013} whether UGR tests with atom interferomters are possible arises.

We treat atomic clocks and atom interferometers in a common framework\,\cite{DiPumpo2021} and discuss UGR tests in atomic clocks.
Via clocks, we define a basic UGR-violation sensitivity and then analyze under which conditions atom interferometers can test UGR.
We show in this article that atom interferometers operated with a single internal state of the atom are fundamentally insensitive to UGR violations.
By introducing superpositions\,\cite{DiPumpo2021,Sinha2011,Zych2011,Loriani2019} of internal states or transitions\,\cite{Roura2020,Ufrecht2020} between them, we identify different classes\,\cite{DiPumpo2021} of atom interferometer schemes which are in principle able to test UGR. 
The main results together with explicit derivations can be found in our previous work \href{https://journals.aps.org/prxquantum/abstract/10.1103/PRXQuantum.2.040333}{PRX Quantum, 2:040333, 2021}.

\section{EEP violations from dilaton fields}
To describe EEP violations, we consider a dilaton field coupling linearly to the Standard Model\,\cite{Damour1994,Damour2010,Damour2012}, which effectively acts as a fifth force.
The Lagrangian of the free part
\begin{equation}
    \label{eq:DilIntLagFree}
    \mathcal{L}_\mathrm{free}=\frac{c^4}{16\pi G}\big[R-2(\nabla\varrho)^2\big]-\frac{1}{4\mu_0}F_{\mu\nu}F^{\mu\nu}-\frac{1}{4}G^{\alpha}_{\mu\nu}G^{\mu\nu}_{\alpha}+\sum_{i=e,u,d}{\bar\psi_i\big[\ii\hbar c\slashed{D}-m_i c^2\big]\psi_i}
\end{equation}
includes a free gravitational contribution with Ricci scalar $R$, Newton's gravitational constant $G$, speed of light $c$, and dilaton field $\varrho$. 
It involves also gauge fields $F_{\mu\nu}$ with vacuum permeability $\mu_0$ and $G_{\mu\nu}$, as well as a matter part with Dirac spinor $\hat{\psi}_i$, Dirac derivative $\slashed{D}$ and mass $m_i$.
The interaction of the dilaton with the gauge fields and the fermions
\begin{equation}
\label{eq:DilIntLag}
        \mathcal{L}_\mathrm{int}=\varrho\left[\frac{d_e}{4\mu_0}F_{\mu\nu}F^{\mu\nu}-\frac{d_g\beta_3}{2 g_3}G^{\alpha}_{\mu\nu}G^{\mu\nu}_{\alpha}\right]-\varrho\sum_{i=e,u,d}{\bar\psi_i(d_{m_i}+\gamma_{m_i}d_g)m_i c^2\psi_i}
\end{equation}
includes unknown coupling constants $d_n$ with $n=m_i,g,e$.
Via these coupling constants, also other constants like masses of composed particles and the fine-structure constant effectively become dilaton dependent\,\cite{Damour1994,Damour2010,Damour2012}.

Hence, analyzing bound systems like atoms in this Standard-Model extension leads to dilaton-dependent energies $E_j(\varrho)$.
These energies can be related to the mass of an atom via the mass-energy relation $E_j(\varrho)=m_j(\varrho)c^2$, where we associate different internal states $j$ with different masses.
Expanding the mass around its Standard-Model value \smash{$m_j(\varrho)\cong (1+\bar\beta_j\varrho)m_j(0)$}, we find the linear coupling coefficient $\bar\beta_j$.
Moreover, the dilaton itself is sourced by Earth, thus depending on gravity via $\varrho=\bar\beta_\mathrm{S} g z/c^2$ including a coupling coefficient $\bar\beta_\mathrm{S}$ for a linear gravitational potential with acceleration $g$.
Other dilaton contributions, e.g. of cosmological origin, are not included for simplicity.
We define the EEP violation parameter
$\beta_j=\bar\beta_\mathrm{S}\bar\beta_j$ which allows the effective replacement $g\rightarrow \left(1+\beta_j\right)g$, introducing a state-dependent gravitational acceleration.

Considering a two-level atom, where $j=a,b$ denotes the excited state $\ket{b}$ and ground state $\ket{a}$, we can identify the mass defect\,\cite{DiPumpo2021,Zych2011,Loriani2019} $\Delta m=m_b(0)-m_a(0)$ and mean mass $m=\left[m_b(0)+m_a(0)\right]/2$, defined via Standard-Model values $m_j(0)$.
The mass defect is related to the internal transition frequency $\hbar\Omega=\Delta m c^2$ of the atom.
To summarize, two perturbative parameters $\Delta m$ and $\beta_j$ have been added to the conventional non-relativistic descriptions.

\section{UGR violations in clocks and quantum clock interferometry}
Based on this underlying model, we will identify a first-quantized Hamiltonian\,\cite{Sonnleitner2018,Schwartz2019} essential for calculating interferometric phases.
After presenting a general perturbative method to calculate such phases, we will consider interferometric clocks and introduce quantum clock interferometry\,\cite{DiPumpo2021,Sinha2011,Zych2011,Loriani2019} based on atom interferometers.
With the help of the differential phases between two clocks, a basic formula for the sensitivity in UGR tests is given.
Then, we will identify UGR-sensitive atom interferometer schemes.
Our treatment follows our previous work published in \href{https://journals.aps.org/prxquantum/abstract/10.1103/PRXQuantum.2.040333}{PRX Quantum, 2:040333, 2021}.

\subsection{Interfering matter waves}
The measured signal $I$ in quantum interference experiment is given by the expectation value of a projection operator $\hat{\Pi}$.
For an initial state $\ket{\Psi_\text{in}} = \ket{\psi_\text{int}}\otimes \ket{\psi_\text{c.m.}}$ where the initial internal state and center-of-mass (c.m.) wave packet are separable, we find
\begin{equation}
    I = \bra{\Psi_\text{in}} \hat{U}^\dagger \hat{\Pi}\hat{U} \ket{\Psi_\text{in}} = \frac{1}{4}  \bra{\psi_\text{c.m.}} (\hat{U}_1^\dagger+\hat{U}_2^\dagger)(\hat{U}_1+\hat{U}_2)\ket{\psi_\text{c.m.}} = \frac{1}{2} (1+C\cos \varphi).
\end{equation}
We define $\big<\hat{U}^{\dagger}_1\hat{U}^{}_2\big>=C \exp({\ii\varphi})$ as the overlap by identifying a superposition $(\hat{U}_1+\hat{U}_2)\ket{\psi_\text{c.m.}}/2$ of two different components.
The specific projection will be discussed below: Depending on the projection, the two components can be related to different spatial branches, internal states, or a combination of them.

We associate the effective Hamiltonian $\hat{H}_j^{(\sigma)}=\hat{H}_0^{(\sigma)}+\hat{\mathcal{H}}_j^{(\sigma)}$ with these components, depending on internal state $j=a,b$ and branch $\sigma=u,l$.
The unperturbed part is given by
\begin{equation}
    \label{eq:HamMassCorr}
    \hat{H}_0^{(\sigma)}=m c^2+\frac{\hat{p}^2}{2m}+m g\hat z-F^{(\sigma)}\hat z+\frac{m\Gamma^2}{2}(\hat{z}-\zeta^{(\sigma)})^2+V_\text{ph}^{(\sigma)}
\end{equation}
for an atom with mass $m$.
Here, we consider a Fermi-Walker expansion of the lab system, leading to a linearly expanded gravitational field with acceleration $g$.
The Hamiltonian includes a branch-dependent constant force $F^{(\sigma)}$, a harmonic potential with frequency $\Gamma$ and time-dependent origin $\zeta^{(\sigma)}(t)$, and a trivial laser phase included in $V_\text{ph}^{(\sigma)}$.
Perturbations to the non-relativistic Hamiltonian are included in 
\begin{equation}
    \label{eq:PerturPot}
    \hat{\mathcal{H}}_j^{(\sigma)}=\lambda_j\frac{\Delta m }{2}\left[c^2-\frac{\hat{p}^2}{2 m^2}+g\hat z\right]+m\beta_j g\hat{z}+\lambda_j\frac{m\Delta\Gamma^2}{4}(\hat{z}-\zeta^{(\sigma)})^2,
\end{equation}
where $\lambda_b=1$ for the excited state and $\lambda_a=-1$ for the ground state.
These perturbations include mass defects $\pm\Delta m$ coupling to the c.m. motion, as well as an EEP-violating factor $\beta_j$.
We also include a state-dependent, perturbative coupling of the harmonic potential via $\lambda_j\Delta\Gamma^2$.

Relying on perturbative methods\,\cite{Ufrecht2019,Ufrecht20202}, we find from this treatment the phase
\begin{equation}
    \label{eq:GeneralPhasePerturb}
    \varphi=\varphi_0-\frac{1}{\hbar}\int\!\text{d}t\,\mathcal{H}_\text{diff}-\frac{1}{2\hbar}\oint\!\text{d}t\,\Big\lbrace\frac{\partial^2\mathcal{H}}{\partial z^2}\left<\hat{z}^2_c\right>+\frac{\partial^2\mathcal{H}}{\partial p^2}\left<\hat{p}^2_c\right> \Big\rbrace,
\end{equation}
where $\mathcal{H}_\text{diff}$ is given by the difference of the classical counterpart of $\hat{\mathcal{H}}_j^{(\sigma)}$, evaluated with the unperturbed trajectories given by Hamilton's equations of motion obtained from the classical counterpart of $\hat{H}_0^{(\sigma)}$.
The explicit form of $\mathcal{H}_\text{diff}$ depends on the branch or the internal state, specified in the following.
The unperturbed phase $\varphi_0$ is generated by $\hat{H}_0^{(\sigma)}$ and the last term describes wave-packet effects. 
These wave-packet effects include the centered position operator $\hat{z}_c(t)$ and momentum operator $\hat{p}_c(t)$ with vanishing expectation values\,\cite{Ufrecht2019,Ufrecht20202}.
They can be calculated from operator-valued equations of motion generated by $\hat{H}_0^{(\sigma)}$, and differ for clocks and atom interferometers.
For the derivatives we find \smash{$\partial^2\mathcal{H}/\partial p^2=-\lambda_j\Delta m/(2m^2)$} and \smash{$\partial^2\mathcal{H}/\partial z^2=\lambda_j m\Delta\Gamma^2/2$}, if the potential is turned on for the whole interferometer.

\subsection{Atomic clocks}
\begin{figure*}
	\centering
	\includegraphics[width=1\textwidth]{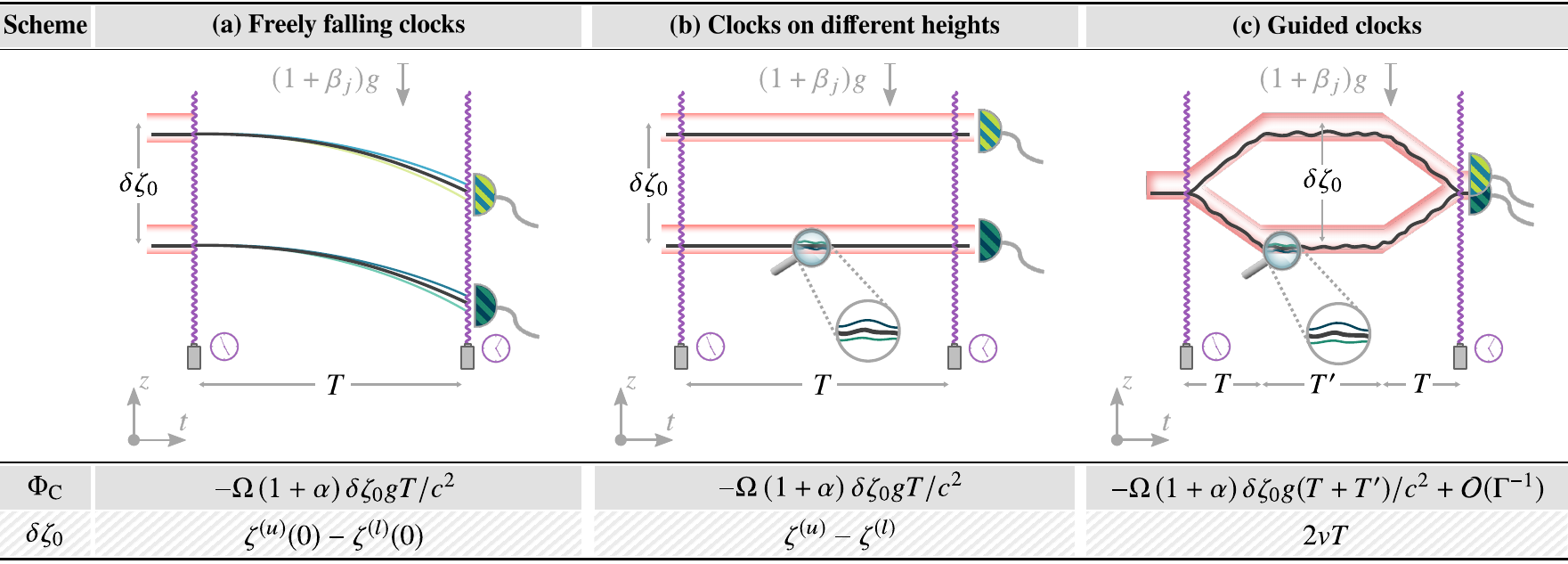}
	\caption{Differential measurements between two clocks in a gravitational field with acceleration $(1+\beta_j)g$: two freely falling clocks (a), two clocks trapped at different but constant heights (b), and two guided clocks being separated and brought back together (c).
	The black lines represent the unperturbed trajectories generated by $H_0^{(\sigma)}$ (surrounded by the perturbed trajectories in color), each trap center is bordered by red potential barriers, and initialization and readout of the clocks is marked by purple pulses.
    The table below shows the differential phase $\Phi_\text{C}$, which yields a UGR sensitivity for all three schemes with height difference $\delta \zeta_0$.
    The Figure was created by and taken from \href{https://journals.aps.org/prxquantum/abstract/10.1103/PRXQuantum.2.040333}{Fabio Di Pumpo et al. PRX Quantum, 2:040333, 2021}, published under a \href{https://creativecommons.org/licenses/by/4.0/}{Creative Commons Attribution 4.0 International license}.
    The Figure was not modified and adopted in its original form created by Fabio Di Pumpo et al.}
\label{fig:DifferentClockGeom}
\end{figure*}

For clocks, we model the readout by a projection onto the superposition $\left(\ket{a}+\ket{b}\right)/2$, which also includes the final pulse.
Hence, we find $\mathcal{H}_\text{diff}=\mathcal{H}_b-\mathcal{H}_a$ for clocks and choose $F^{(\sigma)}=0$, which corresponds to vanishing recoil and no branch-dependent constant force.
With the classical trajectories generated from $H_0^{(\sigma)}$ we can calculate the phase of a single clock.
Comparing two clocks leads to the differential phase
\begin{equation}
    \label{eq:DiffPhaseClocks}
    \Phi_\text{C}=\varphi^{(u)}-\varphi^{(l)}
\end{equation}
between them.
We show the results for three examples\,\cite{DiPumpo2021} in Fig.~\ref{fig:DifferentClockGeom}, where both clocks possess the same transition frequency $\Omega$.
These examples include freely falling clocks where the trap is turned of at $t=0$, clocks on different but constant heights as the prime example for UGR tests, and guided clocks brought to different heights and back together.
We observe that all geometries test UGR with the same differential phase of the form
\begin{equation}
    \label{eq:PhaseIdealClockOnTwoHeights}
    \Phi_\text{C}=\varphi^{(u)}-\varphi^{(l)}=-\Omega\left(1+\alpha\right)\delta\zeta_0 g T/c^2
\end{equation}
where $\zeta_0$ is a height difference, apart from terms suppressed with trap frequency $\Gamma$.
Moreover, we find that UGR tests always involve at least two different internal states.
Here, we defined the parameter $\alpha=m\Delta\beta/\Delta m$, which can be connected to the UFF violation parameter $\Delta\beta$ in the dilaton model.

The centered observables can be calculated as \smash{$\hat{z}_c(t)=\big(\hat{z}-\left<\hat{z}\right>\big)\cos{(\Gamma t)}+\hat{p}\sin{(\Gamma t)}/(m\Gamma)$} and \smash{$\hat{p}_c(t)=\hat{p}\cos{(\Gamma t)}-m\Gamma\big(\hat{z}-\left<\hat{z}\right>\big)\sin{(\Gamma t)}$}.
We assume a vanishing initial momentum $\left<\hat{p}\right>=0$ and vanishing cross terms $\left<\hat{z}\hat{p}+\hat{p}\hat{z}\right>=0$.
Hence, wave-packet effects for clocks turn out to be branch independent and cancel in differential phases.

\subsection{Quantum clock interferometry}
\begin{figure*}
	\centering
	\includegraphics[width=1\textwidth]{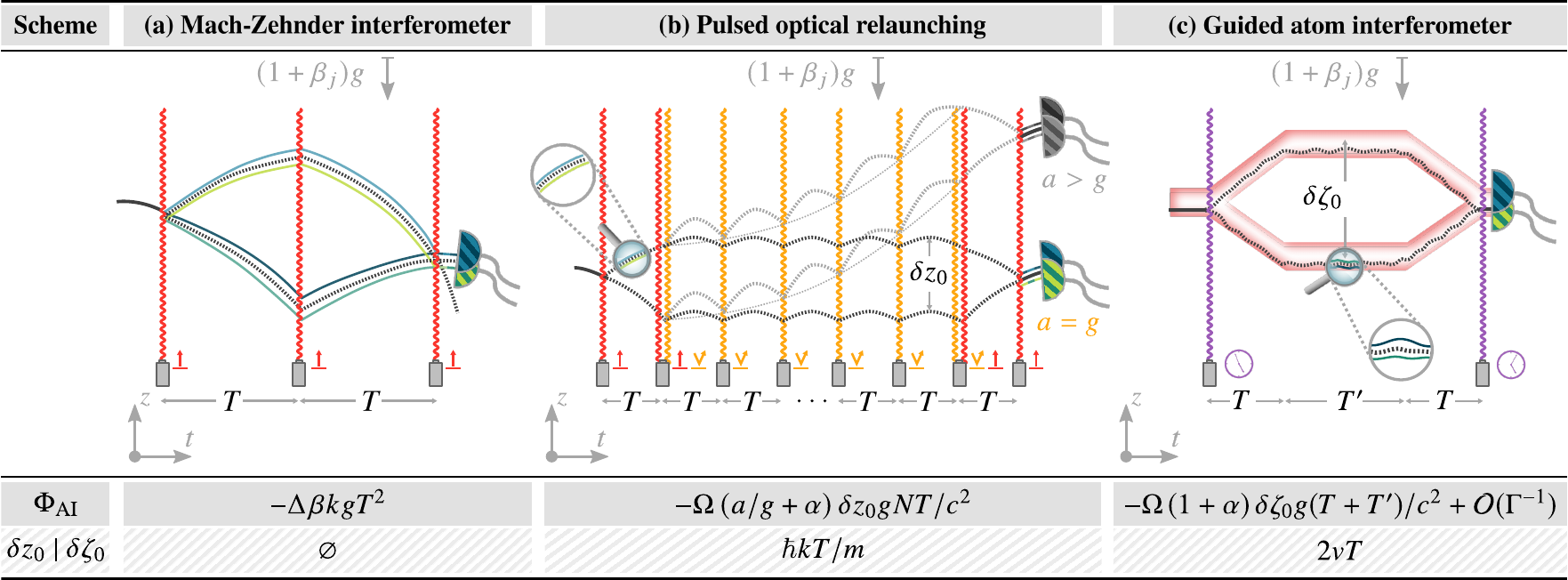}
	\caption{Three examples for quantum clock interferometry in a gravitational field with acceleration $(1+\beta_j)g$:
	a Mach-Zehnder interferometer (a), a scheme including pulsed optical relaunching (b), and an interferometer where two trapping potentials guide the atom to different heights, and back to the initial height (c).
	The Bragg pulses are denoted by red lines, relaunching pulses with effective acceleration $a$ acting on both branches and states alike by yellow ones, and the unperturbed trajectory generated by $H_0^{(\sigma)}$ by dashed lines (surrounded by the perturbed trajectories in color).
	In the table below the differential phases $\Phi_\text{AI}$ are shown.
	While the Mach-Zehnder scheme (a) leads to UFF tests, the pulsed optical relaunching (b) can mimic UGR tests for the specific choice $a=g$.
    The guided scheme (c) is in full analogy to clocks and tests UGR without a specific choice for the acceleration.
    The table also includes height differences $\delta z_0$ or $\delta \zeta_0$.
    The Figure was created by and taken from \href{https://journals.aps.org/prxquantum/abstract/10.1103/PRXQuantum.2.040333}{Fabio Di Pumpo et al. PRX Quantum, 2:040333, 2021}, published under a \href{https://creativecommons.org/licenses/by/4.0/}{Creative Commons Attribution 4.0 International license}.
    The Figure was not modified and adopted in its original form created by Fabio Di Pumpo et al.
    }
\label{fig:DifferentAIGeom}
\end{figure*}
Based on the results for clocks, we analyze which atom interferometers without internal transitions (Bragg-type) display the same UGR-sensitive phase.
For Bragg-type atom interferometers\,\cite{Giese2015} one can project onto a specific momentum (range) at the end of the experiment, defining the projection $\hat{\Pi}$.
Hence, we find $\mathcal{H}_{\text{diff}}=\mathcal{H}^{(u)}-\mathcal{H}^{(l)}$ for Bragg-type atom interferometers.
For light-pulse atom interferometers, we have $\Gamma=\Delta\Gamma=0$.
With the classical trajectories generated from $H_0^{(\sigma)}$ we can calculate the phase of a single atom interferometer, analogously to clocks.
Comparing phases of two atom interferometers operated in a different internal state, performed simultaneously or sequentially, leads to the concept of quantum clock interferometry\,\cite{DiPumpo2021,Sinha2011,Zych2011,Loriani2019} and yields the differential phase
\begin{equation}
    \label{eq:DiffPhaseAIs}
    \Phi_\text{AI}=\varphi_b-\varphi_a.
\end{equation}
We show in Fig.~\ref{fig:DifferentAIGeom} three examples\,\cite{DiPumpo2021} for quantum clock interferometry.
Whereas the Mach-Zehnder interferometer serves as a prime example for such schemes, it does not provide a UGR test but rather a UFF test, since it is directly proportional to $\Delta\beta = \beta_b-\beta_a$ and has the form of a Null test. 
Introducing relaunch pulses with effective acceleration $a$ acting on both branches alike, the differential phase includes a term $a/g$.
For $a$ exactly tuned to have the same value as $g$, this phase mimics UGR tests.
Contrarily, by using guided schemes where the trapping potential is not turned off, quantum clock interferometry leads automatically to UGR tests, in complete analogy to clocks.
We observe that for UGR tests with atom interferometers two internal states are involved.
Hence, a single internal state cannot be sufficient, as the transition frequency $\Omega$ can only be obtained as a prefactor in the differential phase by involving two internal states.

The centered observables $\hat{z}_c(t)=\hat{z}-\left<\hat{z}\right>+(\hat{p}-\left<\hat{p}\right>)t /m$ and $\hat{p}_c=\hat{p}-\left<\hat{p}\right>$ for Bragg-type atom interferometers, as well as the derivatives \smash{$\partial^2\mathcal{H}/\partial p^2$} and \smash{$\partial^2\mathcal{H}/\partial z^2$}, are branch independent.
Thus, no wave-packet effects arise.

\section{UGR violations via internal transitions}
\begin{figure*}[h]
	\centering
	\includegraphics[width=1.0\textwidth]{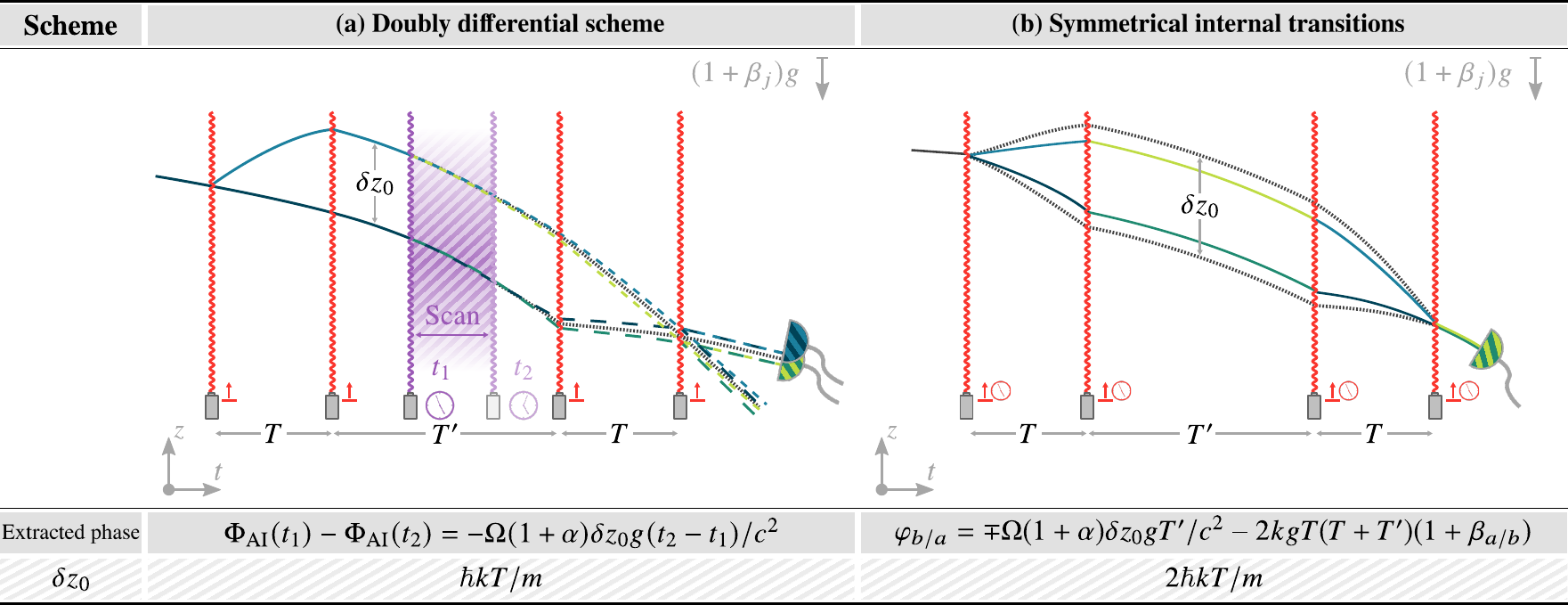}
	\caption{Two examples for UGR tests with atom interferometers relying on internal transitions during the sequence in a gravitational field with acceleration $(1+\beta_j)g$:
	a Ramsey-Bord\'{e}-like geometry operated as a doubly differential scheme (a) and a scheme relying on sequential internal transitions (b).
	The table below the figures shows the phases and height differences $\delta z_0$.
	The unperturbed trajectories generated by a generalization of $H_0^{(\sigma)}$ are denoted by dashed lines (surrounded by the perturbed trajectories in color), while red and purple lines represent Bragg pulses and recoilless pulses, the latter brings an atom into an internal superposition.
	After applying two Bragg pulses opening the interferometer in the first scheme (a), the recoilless pulse at $t_1$ initializes a clock while the atom drops in superposition of a height difference $\delta z_0$.
	Since the differential phase $\Phi_\text{AI}(t_1)$ depends on that time, a UGR sensitivity can be obtained by performing the scheme a second time but with another initialization time $t_2$, leading to a redshift measurement in the purple shaded area.
	Contrarily, the second scheme (b) uses pulses which both transfer momentum and change the internal state.
	Via these symmetrical internal transitions, the phase $\varphi_{b/a}$ becomes sensitive to UGR violations without a differential scheme, which, however, should be performed also for this example to remove cloaking effects.
	The Figure was created by and taken from \href{https://journals.aps.org/prxquantum/abstract/10.1103/PRXQuantum.2.040333}{Fabio Di Pumpo et al. PRX Quantum, 2:040333, 2021}, published under a \href{https://creativecommons.org/licenses/by/4.0/}{Creative Commons Attribution 4.0 International license}.
    The Figure was not modified and adopted in its original form created by Fabio Di Pumpo et al.}
\label{fig:DifferentAIGeom2}
\end{figure*}

So far we considered solely internal superpositions but can generalize the treatment also to internal transitions\,\cite{Roura2020,Ufrecht2020}.
In this case, we find from a similar formalism UGR-sensitive phases.
We show two examples\,\cite{Roura2020,Ufrecht2020} for this class of geometries in Fig.~\ref{fig:DifferentAIGeom2}.
Via these transitions, two internal states get involved.
Again, UGR-sensitive phases can be extracted\,\cite{DiPumpo2021}, without necessarily involving an internal superposition.
Additionally, the second scheme also includes UFF-sensitive contributions, which can be separated by varying the middle time segment.

\section{Conclusion}
We derived a UGR-violating expression for atomic clocks based on a dilaton model.
By treating clocks and atom interferometers on the same footing, we showed that UGR tests have to involve two internal states.
Finally, we analyzed which atom-interferometric schemes can test UGR and found that this sensitivity can be achieved either by mimicking or guiding schemes for quantum clock interferometry, or by driving internal transitions during the interferometer.

\newpage
\section*{Acknowledgments}
I thank C. Ufrecht, A. Friedrich, E. Giese, W. P. Schleich, and W. G. Unruh for the fruitful collaboration, culminating in the publication \href{https://journals.aps.org/prxquantum/abstract/10.1103/PRXQuantum.2.040333}{Fabio Di Pumpo et al. PRX Quantum, 2:040333, 2021}.
The projects ``Metrology with interfering Unruh-DeWitt detectors'' (MIUnD) and ``Building composite particles from quantum field theory on dilaton gravity'' (BOnD) are funded by the Carl Zeiss Foundation (Carl-Zeiss-Stiftung).
The work of IQ\textsuperscript{ST} is financially supported by the Ministry of Science, Research and Art Baden-W\"urttemberg (Ministerium f\"ur Wissenschaft, Forschung und Kunst Baden-W\"urttemberg).
The QUANTUS project is supported by the German Aerospace Center (Deutsches Zentrum f\"ur Luft- und Raumfahrt, DLR) with funds provided by the Federal Ministry for Economic Affairs and Climate Action (Bundesministerium f\"ur Wirtschaft und Klimaschutz, BMWK) due to an enactment of the German Bundestag under Grant Nos. 50WM1956 (QUANTUS V) and 50WM2250D (QUANTUS+).

\section*{References}
\bibliography{FabioDiPumpo}

\end{document}